\documentclass[12pt,letterpaper]{article}
\usepackage{color}
\usepackage{amsmath}    
\usepackage{amssymb}
\usepackage{graphicx} 
\DeclareGraphicsExtensions{.png,.jpg,.gif,.ps,.eps}
\usepackage{float}
\usepackage{authblk}
\usepackage[a4paper]{geometry}
\usepackage[final]{pdfpages}
\usepackage{epstopdf}

\begin{document}     

\title{Vegetation pattern formation in a sinuous free-scale
landscape}

\author[1]{Rub\'en Mart\'inez D.} 
\author[2]{Andrea Montiel P.}
\author[3]{J. Fernando Rojas}
\affil[1,2,3]{Facultad de Ciencias F\'isico Matem\'aticas.\\ Benemerita Universidad Aut\'onoma de Puebla,\\Av. San Claudio y 18 sur, Ciudad Universitaria, Col. San Manuel.\\ C. P. 72570. Puebla, M\'exico.}
\affil[3]{Centro Multidisciplinario de Modelaci\'on Matem\'atica y Computacional. Benem\'erita Universidad Aut\'onoma de Puebla.}
%\author[1,2,3]{Victor Buend{\'\i}a}
%\author[1]{Pablo Villegas}
%\author[4]{Serena di Santo} 
%\author[2,5]{Alessandro Vezzani} 
%\author[2,3]{Raffaella Burioni}
%\author[1,2]{Miguel A. Mu\~noz}
  
  %\affil[1]{Departamento de Electromagnetismo y F{\'\i}sica de la}
  %Materia e Instituto Carlos I de F{\'\i}sica Te\'orica y
  %Computacional. Universidad de Granada.  E-18071, Granada, Spain}
  %\affil[2]{Dipartimento di Matematica, Fisica e Informatica,
  %Universit\`a di Parma, via G.P. Usberti, 7/A - 43124, Parma, Italy}
  %\affil[3]{INFN, Gruppo Collegato di Parma, via G.P. Usberti, 7/A -
  %43124, Parma, Italy}
  %\affil[4]{Scuola Internazionale Superiore
  %di Studi Avanzati, via Bonomea, 265 - 34136 Trieste, Italy.}
  %\affil[5]{IMEM-CNR, Parco Area delle Scienze 37/A - 43124 Parma, Italy}

\date{}  
\maketitle  
  
\begin{abstract}

The original Hardenberg's model of biomass patterns in arid and semi-arid regions is revisited to extend it to more general -non flat- regions. It is proposed a technique to study these more generalized (non-flat) regions using both a conservation criterion and a explicit spatial dependant function $\nu (x)$. In this paper a study of dynamical stability around sistem's fixed points made. Under the idea of predictibility via air images a fitted relationship among dynamical variables at stable fixed points is stablished. Also, is presented a discrete version of the model, in the form of Cellular Automata thechniques, that allows to neglect the spatial scale and reproduces realistic stable spatial patterns. \\
    
\end{abstract}

%\begin{keyword}
%EEG \sep Epilepsy \sep Feigenbaum graphs  \sep visibility graph
%% keywords here, in the form: keyword \sep keyword

%% MSC codes here, in the form: \MSC code \sep code
%% or \MSC[2008] code \sep code (2000 is the default)

%\end{keyword}

%% main text
\section{Introduction}
\label{S:1}
The structure and dynamics of ecosystems are extremely complex. The interactions among the multiple species and its interactions with their landscape, climate conditions and human behaviour are quite complex. There exists a lot of ways in which these interactions can be present, in particular in the animal-vegetation interactions case, where they can change the behavior and population of both vegetation and animal species. In recent ecology studies {\cite{aboyer}} has been shown that the inhomegeneities in food resources modifies the foraging walk dynamics of herbivore species. This inhomegeneities can be modeled in a spatial sense or in terms of their importance or size as food resources.\\
  Not only the walking dynamics but the social behavior of animal population could be modified, as the Ramos-Fern\'andez et. al. {\cite{aramos2006}} model suggests. The subgroups formation in hervibore mammals create a social network which structure depends on the conjugation of animal's memory and the food patches size distribution. Furthermore, animal and vegetation species can be thinked as ``ecosystem's engineers'', i. e., key species that modulate the introducction, permanence or extintion of another ones in the ecosystem {\cite{agilad}}.\\ %\begin{flushleft}

  In the case of ground vegetation, these interactions could be the responsibles of the wide variety of propagation, stablishment and survive strategies. For example, vegetation propagation can be done via seed dispersion, by hervibores or by wind, and by asexual reproduction, in which a new individual is conected to the parental plant. The latter is important in low humidity regions because the conection with the parental plant provides the new one a certain independence of landscape characteristics, such as surface watter or soil nutrient availability. Also, recent work of Thompson {\cite{athompson2008}} shows that seed dispersal mechanism changes the spatial configuration of vegetation, destabilizing the regular patterns observed in diffusion-based models. In the other hand, propagation via seeds gives the vegetation another way to survive, as the seeds can be lattent, in some cases, for long time periods between seed dispersal and germination, and, furthermore, the  seeds can survive low watter or fire regimes. In the latter case, there are   seeds that survives high temperatures regimes, around $90^{\circ} C$, as   observed in {\em{Bombax tomentosum, Bowdicha major, Brosimum   gaudichaudii}} and other species. Indeed, another seed species seem to be   favoured by this situation, for example {\em{M. pubescens}} and {\em{B. major}} {\cite{anunez}}.
  
  As above, interactions between vegetation and soil features are diverse. As an example, it is observed that the flower and fruit generation in some species are sinchronized with dry season in savhannas, wich is an evidence of deep watter resources available for some species. Also, soil nutrients distribution could be an important feature in the species spatial configuration, as in savhannas where is observed a grassland continous matrix ocasionally interrupted by low density tree patches.
  
  Also, vegetation dynamics can modulate their landscape and the pressence or absence of animal populations. Vegetation is traditionaly used for soil recovering, as it prevents erosion via its roots, for soil structure improvement, as it provides physical support and organic matter addition, and acelerates nutrient fixation process.
  
  Another interesting feature of vegetation dynamics is the pattern formation process which is supposed to be the result of cooperation and competition mechanisms between vegetation patches in a limited resources landscape {\cite{ahardenberg}}, relationships between bacteria and vegetation species {\cite{agilad}} and stochastic phenomena {\cite{ashnerb}}. The Figure below, shows some vegetation patterns observed in arid and semiarid lands.
%\end{flushleft}

\begin{figure}[H]
  \centering
  \includegraphics[width=5in]{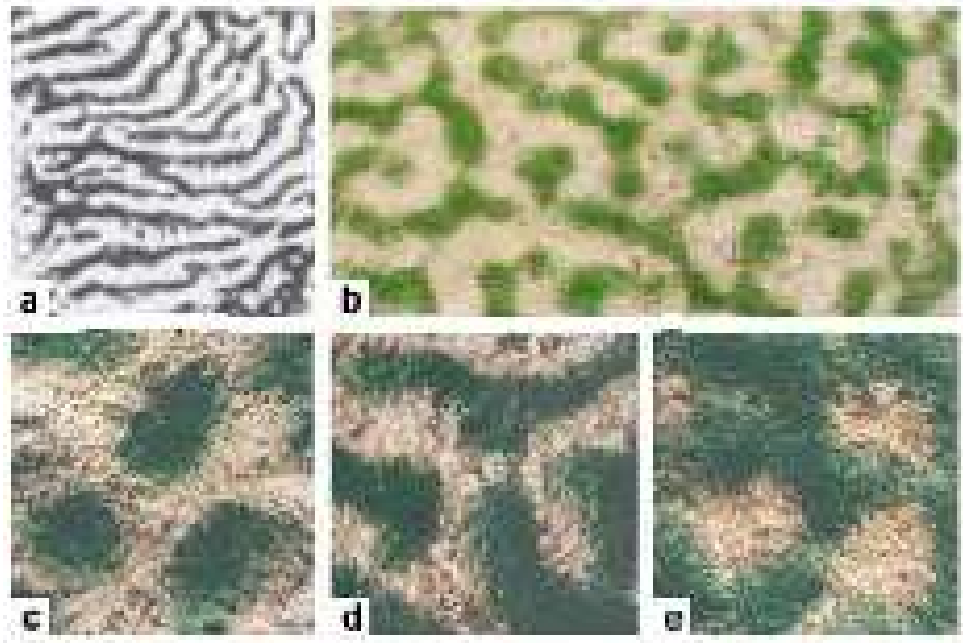}
  \caption{\label{originalH}Original image taken of Hardenberg's paper {\cite{ahardenberg}}. The scales in the size of biomass patterns are quite different: from centimeters (for patterns $c$, $d$ and $e$) to tens of meters (patterns $a$ and $b$)}.
  \label{fig1}
\end{figure}

%\begin{flushleft}
  As sketched above, these interactions, and several another, are
  extremely complicated to model, but it is important to account them in the
  search for more complete and accurate descriptions of ecosystems dynamics.
  It is necessary to consider both animal species in mutual interactions and
  the vegetal species vinculated with the animal behavior. Mineral resources,
  weather, humedity, chemical and physical properties of soil and different
  type of relations between species are topics of importance in this theme.
  However, a model that consider all the interactions mentioned above will be
  quite complicated.
  
\begin{figure}[H]
  \centering
  \includegraphics[width=5in]{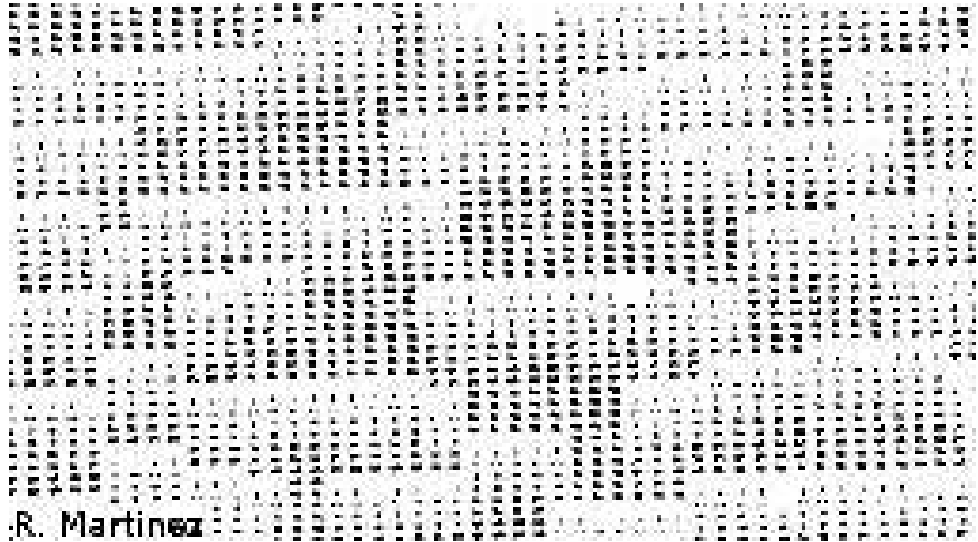}
  \caption{Shigamare dynamic patterns of propagating wave front. Dead trees are represented by darkest cells. }
  \label{fig2}
\end{figure}

  On the other hand, there exists simplified models and reports
  that enables the comprehension of some interesting and important aspects of
  this system. These models are developed using different techniques, such as
  Partial Differential Equations (PDE), Reaction-Diffusion Systems (RDS),
  Cellular Automata (CA), etc. For example, the work cited at Sol\'e's book
  {\cite{lsole}} consists of a simple CA model to describe the observed
  behavior in some american and japanesse rainforest regions, called Shigamare
  effect. These regions presents a kind of propagating ``wave front'' of dead
  trees over the entire forest. In Sol\'e's model, each cell of the CA
  represents, via its value, a tree of some size (or age). The temporal
  evolution of the tree size depends on the average size of its neighborhood
  and on the wind direction. These assumptions are the responsibles of the
  dynamic patterns obtained with this model shown in Figure \ref{fig2}.
%\end{flushleft}

\begin{figure}[H]
  \centering
  \includegraphics[width=4in]{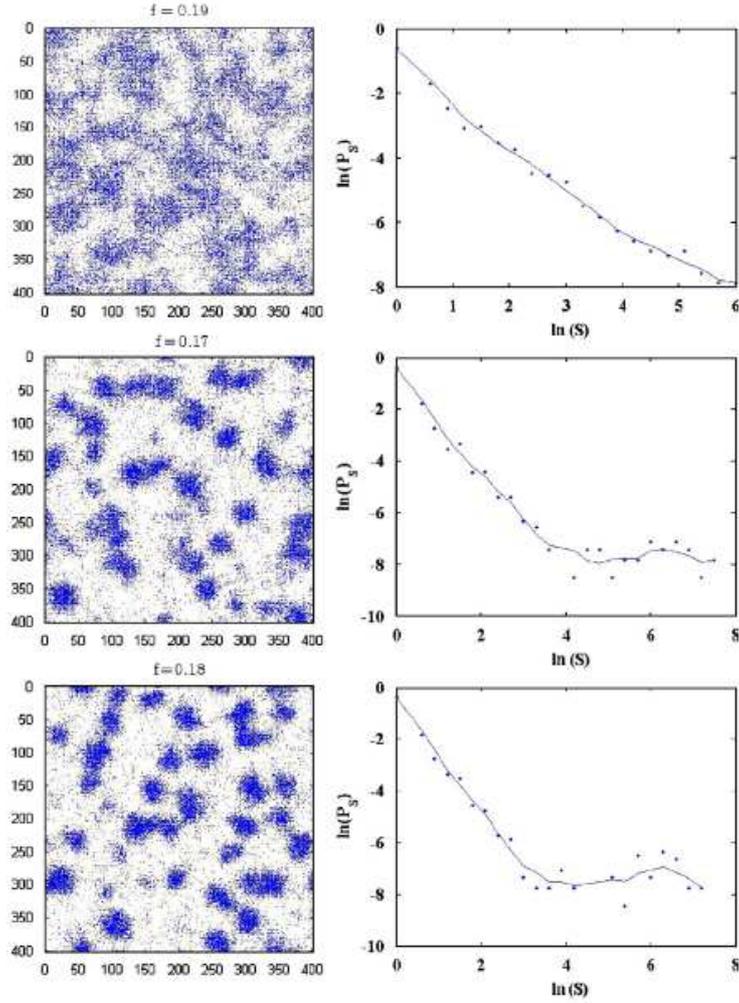}
  \caption{Spatial vegetation patterns and patches size distributions obtained v\'{\i}a Shnerb's model. Note that in the second and third distributions the L\'evy-like regime is lost.}.
  \label{fig3}
\end{figure}

%\begin{flushleft}
  Another approach to vegetation dynamics was made by Manor and
  Shnerb for the vegetation patterns observed in arid and semi-arid regions
  near Jerusalem and Nigeria {\cite{amanor2008}}. This model consists on two
  diferential equations, one for biomass and another for soil water density
  which are modeled as discrete local variables as values in a finite lattice
  of lenght $L$. They consider a positive feedback mechanism related with two
  process of vegetation mortality: below a threshold size the new vegetation
  shrubs wont survive and natural death. Both mechanisms are stochastic and
  obey two kinds of probabilities which depends on the distance between shrubs
  and the threshold size. The results of this model are presented in terms of
  stable vegetation patterns and L\'evy-like patches size distributions as
  shown in Figure \ref{fig3}.
%\end{flushleft}

%\begin{flushleft}
The aim of this work is, in one hand, to make a detailed analysis of Hardenberg's model related with its fixed points, stability around them and to study statistical distributions of biomass densities. The other part of this study is related with the possibility that, under random initial conditions in a uniform grid, this model can generate the biomass patterns in a non-flat region.\\
  
In section 2 the stability analysis of the Hardenberg's model is done, showing that, close to the non obvious fixed point, the system is asymptotically stable and the biomass density tends to a value that depends on the $p$ parameter. Also, here is proposed a discrete version of the original model that conduces to the neglecting posibility of the spatial scale and two ways to generalize the vegetation patterns in a non-flat region. Section $3$ is devoted to the results obtained in terms of vegetation patterns formation and the biomass density dristributions. Finally, a discussion over the results is in Section 4.\\

 \section{Hardenberg's model}

  {\noindent}The original model of Hardenberg {\cite{ahardenberg}} consists of
  a pair of dynamical equations that involves terms related with diffusion for
  both green biomass (plants) density $n$ and soil humidity $w$. The coupled
  equations are
%\end{flushleft}

\begin{equation}
  \frac{\partial n}{\partial t} = \frac{\gamma w}{1 + \sigma w} n - n^2 - \mu
  n + \nabla^2 n \label{hard1}
\end{equation}
and
\begin{equation}
  \frac{\partial w}{\partial t} = p - (1 - \rho n) w - n w^2 + \delta \nabla^2
  (w - \alpha n) + v \frac{\partial (w - \beta n)}{\partial x}  \label{hard2}
\end{equation}

%\begin{flushleft}
The first term in Eq. (\ref{h1}) accounts for the biomass growth rate with the present local biomass and a coefficient that accounts for a limited influence of the pressence of humidity. The second term is vinculated with finitude of resources, as in logistic model of populations growth {\cite{bstrogatz}}. Quantity $- \mu n$ represents dead of vegetation, due both to herbivores and natural mortality. In second equation, $p$ is a source term for the humidity, as an annual rain average. Second and third terms are related to watter loss due evaporation, which is reduced in the presence of vegetation by the factor $(1 - \rho n)$, and due to watter uptake by the vegetation's roots. In both equations, the spatial terms, $\nabla^2 n$ and $\delta \nabla^2 (w - \alpha n)$, models the diffusion of biomass and humidity in space, with the latter reduced by the prescence of biomass. In the last term in Eq. (\ref{h2}), $v$ is the watter runoff velocity towards direction $x$, where a hill slope is assumed, and the partial derivate models the humidity diffusion in this direction. \\
  
\begin{figure}[H]
  \centering
  \includegraphics{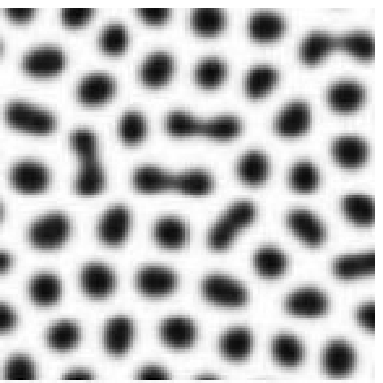}
  \includegraphics{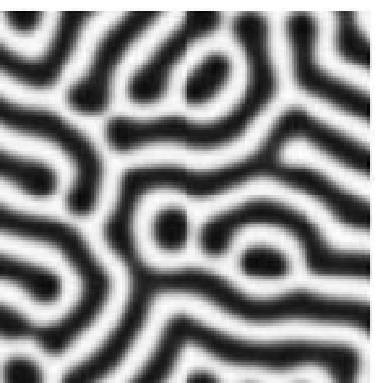}
  \includegraphics{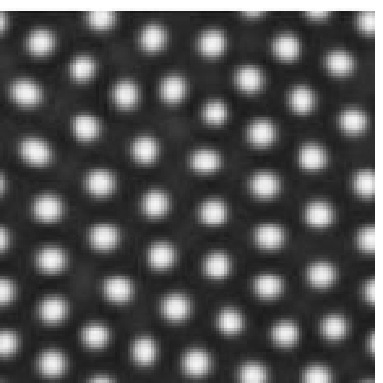}
  
  \includegraphics{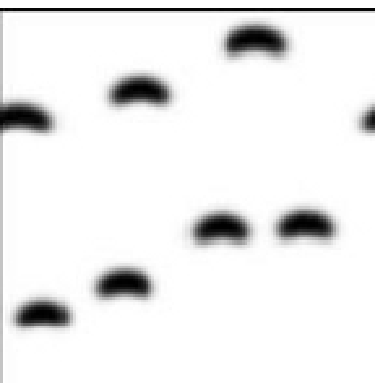}
  \includegraphics{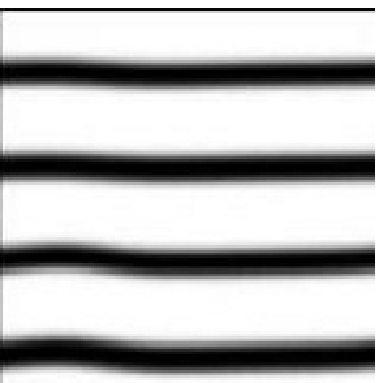}
  \includegraphics{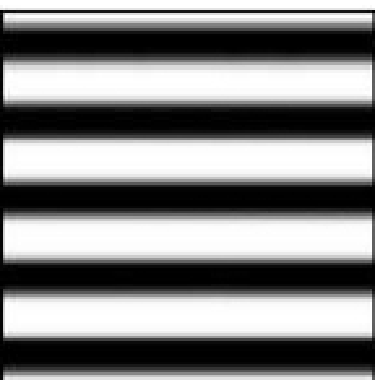}
  \caption{Vegetation patterns resulting from Hardenberg's model. Note the linear behaviour of banded patterns. Reprinted figures with permission from: J. von Hardenberg, E. Meron, M. Shachak and Y. Zarmi. Physical Review Letters 87, 198101-1, 2001. Copyright(2001) by the American Physical Society. http://prl.aps.org/abstract/PRL/v87/i19/e198101}
   \label{fig4}
\end{figure}

With these assumptions, Hardenberg's model show a good aproximation to real vegetation patterns (Figure \ref{fig1}) as can be seen in Figure \ref{fig4}. \\
%\end{flushleft}

\subsection{Stability analysis}

%\begin{flushleft}
  {\noindent}In order to know the possible correlations between local biomass
  density, humidity and the control parameter $p$ -that defines the model
  dynamic behavior- in this section we made a linear stability analysis
  {\cite{bstrogatz}}, which enable us to find all the system's fixed points in
  terms of $p$.

%\end{flushleft}

If we define
\[ F_n (n, w) = \frac{\gamma w}{1 + \sigma w} n - n^2 - \mu n \]
and
\[ F_w (n, w) = p - (1 - \rho n) w - n w^2, \]
fixed points $(n^{\star}, w^{\star})$ of the system

\begin{eqnarray*}
  \frac{\partial n}{\partial t} = F_n (n, w) & ; & \frac{\partial w}{\partial
  t} = F_w (n, w)
\end{eqnarray*}

%\begin{flushleft}
  {\noindent}are obtained for each value of $p$ solving $F_n (n^{\star},
  w^{\star}) = F_w (n^{\star}, w^{\star}) = 0$. So, with the aim of
  determining the stability of Hardenberg's model, we evaluate the aproximate
  numerical values of the fixed points for a series of values of $p$
  parameter, some of them shown in Table \ref{tpfijosH}.
%\end{flushleft}

\begin{center}
  %\float{h}{small}{table}{
  \begin{table}
  \centering
  \begin{tabular}{|c|c|c|c|c|c|c|c|c|}
    \hline
    $p$ & 0.20 & 0.30 & 0.40 & 0.50 & 0.60 & 0.70 & 0.80 & 0.90\\
    \hline
    $n^{\star}$ & 0.05654 & 0.17379 & 0.26110 & 0.32059 & 0.36225 & 0.39308 &
    0.41704 & 0.43639\\
    \hline
    $w^{\star}$ & 0.21566 & 0.37306 & 0.53477 & 0.67871 & 0.80276 & 0.91093 &
    {\em{1.00703}} & {\em{1.09387}}\\
    \hline
    
  \end{tabular}
  \caption{Fixed Points obtained numerically for some values of $p$. The other parameters are \ $\delta = 100, \alpha = 3, \beta = 3, \sigma = 1.6, \gamma = 1.6, \mu = 0.2, \rho = 1.5$ in all the calculations.}
  \label{tpfijosH}
  \end{table}
\end{center}

%\begin{flushleft}
 In making the linear stability analysis we define $u_n = n - n^{\star}$ and $u_w = w - w^{\star}$, whose numeric values are small, and construct the jacobian matrix associated with the dynamical system. Then, the original model in a small region around fixed points can be wroten as:
%\end{flushleft}

\[ \left( \begin{array}{l}
     \frac{\partial u_n}{\partial t}\\
     \\
     \frac{\partial u_w}{\partial t}
   \end{array} \right) = \left( \begin{array}{ll}
     \left. \frac{\partial F_n}{\partial n} \right|_{P^{\star}} & \left.
     \frac{\partial F_n}{\partial w} \right|_{P^{\star}}\\
     & \\
     \left. \frac{\partial F_w}{\partial n} \right|_{P^{\star}} & \left.
     \frac{\partial F_w}{\partial w} \right|_{P^{\star}}
   \end{array} \right) \left( \begin{array}{l}
     u_n\\
     \\
     u_w
   \end{array} \right) . \]

%\begin{flushleft}
  {\noindent}It is found that all eigenvalues of the Jacobian matrix are
  negative and, then, all fixed points are attractor points: the model is
  stable near of these points.
  
\begin{figure}[H]
  \centering
  
  \includegraphics[width=3in]{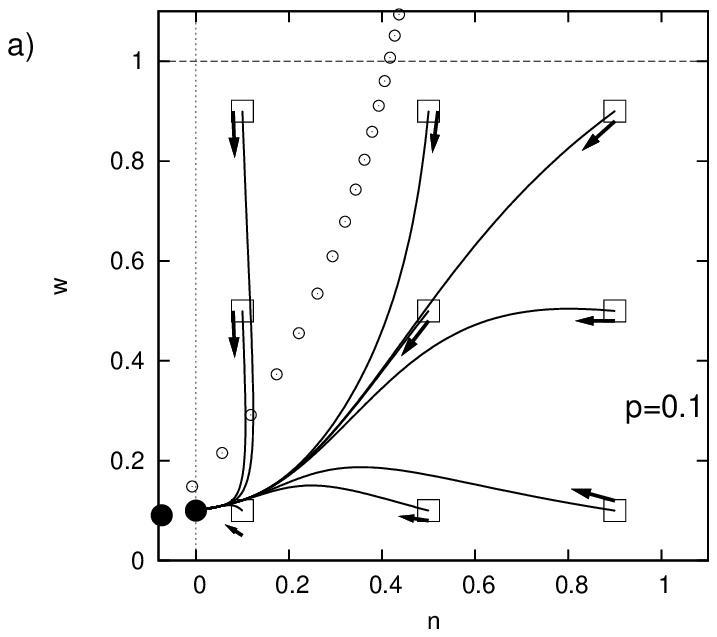}\includegraphics[width=3in]{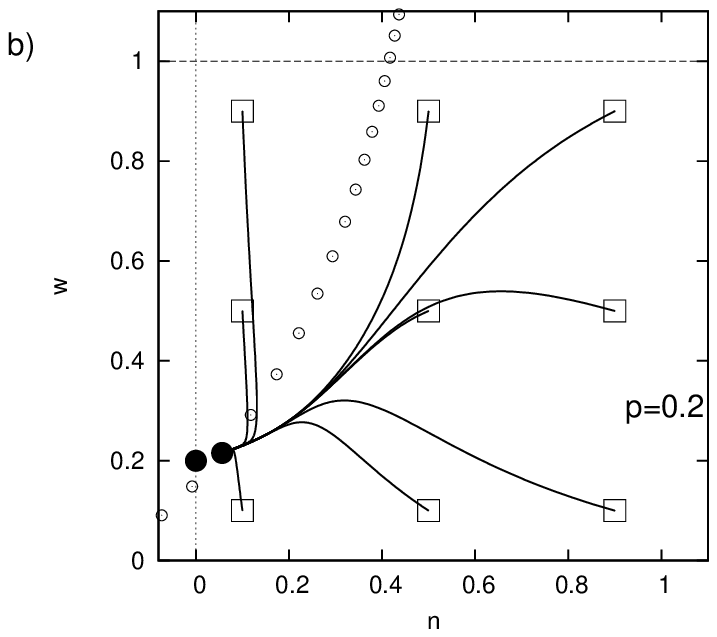}
  
  \includegraphics[width=3in]{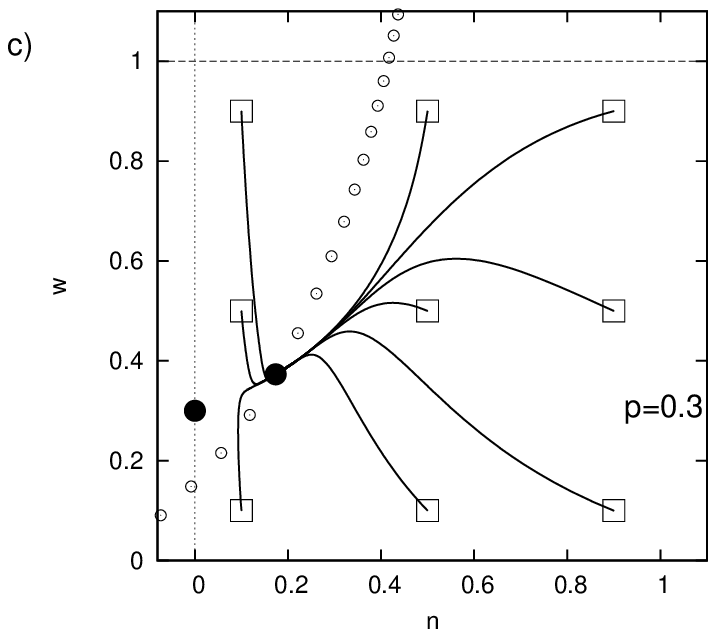}\includegraphics[width=3in]{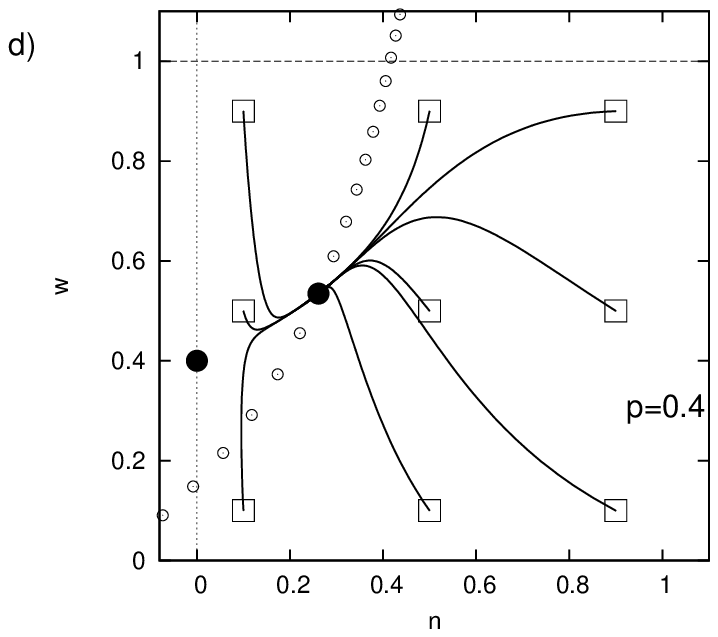}
  
  \includegraphics[width=3in]{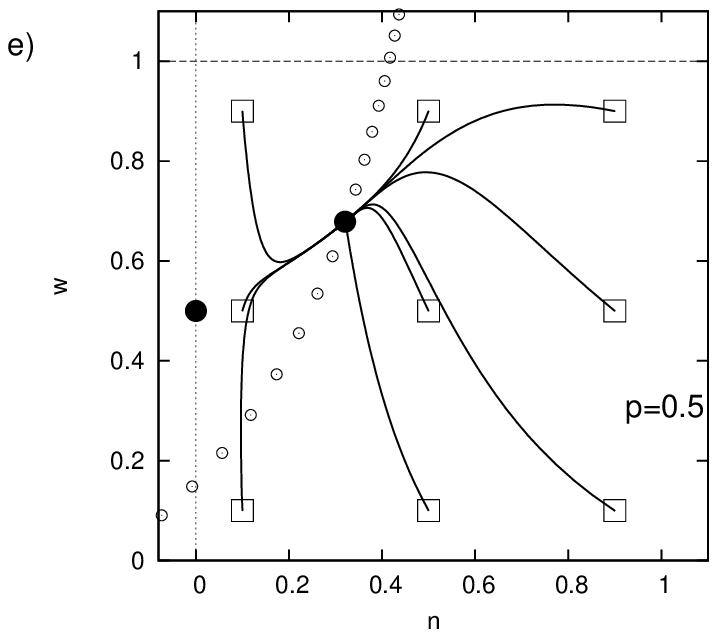}\includegraphics[width=3in]{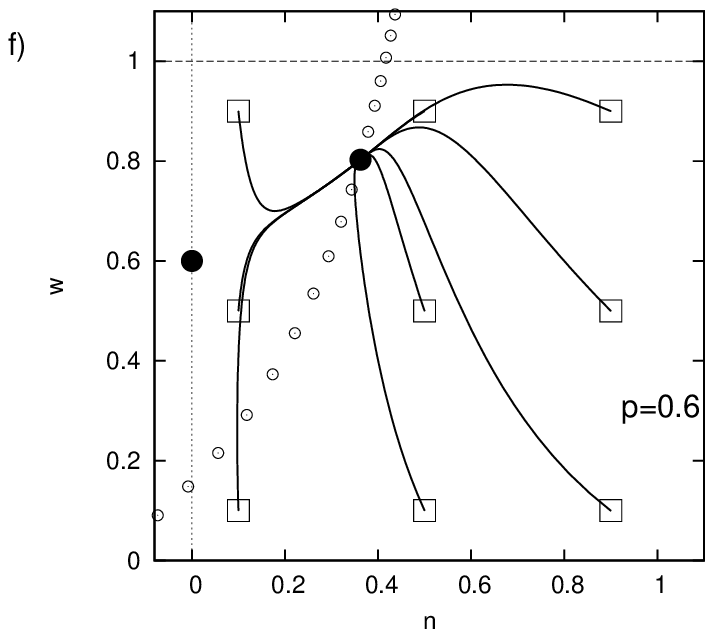}
  \caption{All the initial conditions for $n$ and $w$ in the phase space for $p = 0.1, 0.2, \ldots, 0.6$ leads the system toward its fixed points. Square symbols represents initial conditions. Filled black dots are the fixed points for a particular $p$ and white circles shows a sequence of fixed points for $p = 0.1, \ldots, 0.8$. For each $p$ value the obvious fixed point $(0, p)$ is on the left of image, except in a) where $n^{\star} < 0$. All the images above were performed for 300,000 discrete time steps. Arrows in a) show the flow direction.}
  \label{puntosfijosH}
\end{figure}

  {\noindent}In correspondence, neglecting the explicit spatial terms in the
  original equations (\ref{hard1}) and (\ref{hard2}), a rasoneable set of
  initial conditions is selected with the aim of verify that the $[0, 1]
  \times [0, 1]$ region of phase space is contained in an attraction basin
  {\cite{bstrogatz}} and to evaluate the dynamical evolution of the variables.
  Some fixed points are presented in Figure \ref{puntosfijosH} with small
  white circles and the particular ones, associated with each $p$ value, as an
  out-of-axis big black dot{\footnote{The {\em{obvious}} fixed point $(0,
  p)$ appears on the $n = 0$ vertical axis.}}.
%\end{flushleft}

{\color{black} }

%\begin{flushleft}
  {\noindent}It is clear, from Figure \ref{puntosfijosH} a), that very small
  values in the $p$ parameter has no sense in the dynamics of $n$ and $w$. The
  fixed poin for $p = 0.1$, for example, corresponds to a negative value of
  the biomass density $n$, and appears at the left of the $(0, p)$ fixed
  point.
  
\begin{figure}[H]
  \centering
  \includegraphics[width=4in]{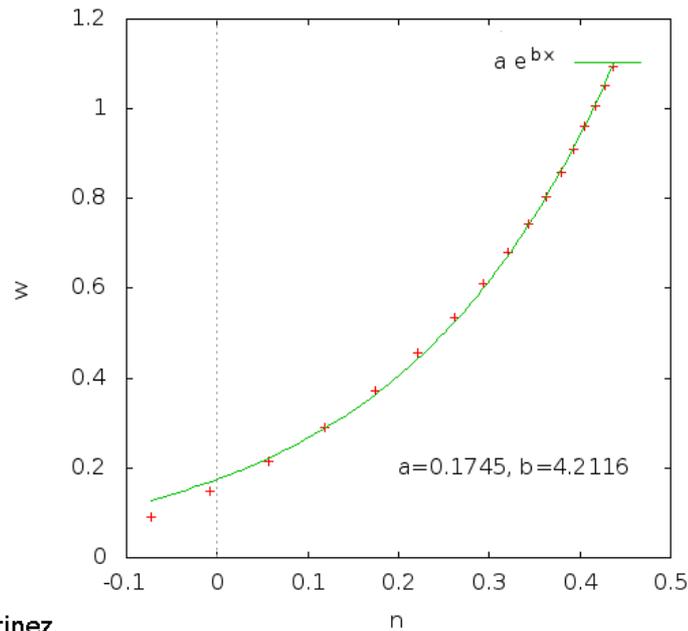}
  \caption{\label{fixedpoints}Fitted fixed points in phase space of
  Hardenberg's model for diferent $p$ values.}
\end{figure}

  {\noindent}The shown sequence of fixed points (small white circles in Figure
  \ref{puntosfijosH}) appears to be a functional relationship between the $w$
  and $n$ variables that could be used to determine humidity density from
  biomass density values. In Figure \ref{fixedpoints} is shown an exponential
  curve fitted over the fixed points which represents the functional
  relationship between biomass and humidity fixed points
%\end{flushleft}

.

\subsection{Discretization}\label{discrete}

%\begin{flushleft}
  {\noindent}Note that the nonlinear character of equations (\ref{hard1}) and
  (\ref{hard2}) implies the numeric solution of the first order temporal part
  by means o\label{sinu}In a similar grid than Figure \ref{flat512patterns}
  using the smoothing function $\nu (x)$ in Eq. (\ref{nudex}). The pictures
  above were rotated $180^{\circ}$ to show the correspondence with those in
  Figure \ref{sisalio}f an appropiate numeric approximation, using some
  general numeric algorithm such as Euler or Runge-Kutta methods.
  
  {\noindent}In the spatial case, the discretizaci\'on via finite differences
  of the diffusion (laplacian) operator implies that the neighboring cells to
  each one modifies the local state of last. If the grid has a size $h$, or
  the cells $h^2$, in the lattice then, for a continuous $C^2$ class function,
  must be true:
%\end{flushleft}
\begin{eqnarray*}
  \nabla^2 u (x, y, t) & \simeq & \frac{1}{h^2} [u (x + h, y, t) + u (x - h,
  y, t) +
\end{eqnarray*}
\begin{equation}
  + u (x, y + h, t) + u (x, y - h, t) - 4 u (x, y, t)] . \label{laplaciano}
\end{equation}

%\begin{flushleft}
  {\noindent}In order to observe how the model's dynamics can develop the
  desired spatial patterns, it is important that these patterns can be
  obtained from a local random initial state. However, in doing so, the
  function $u$ does not longer belong to $C^2$ class functions, and the
  algorithm stability disappears. So that, in order to stablish random values
  at $t = 0$, we redefine the discrete Laplace's operator as:
%\end{flushleft}
\begin{equation}
  \nabla^2 u_{i, j}^t = u^t_{i, j + 1} + u^t_{i, j - 1} + u^t_{i + 1, j} +
  u^t_{i - 1, j} - 4 u^t_{i, j}  \label{lapdiscreto},
\end{equation}

%\begin{flushleft}
  {\noindent}where the symbol $u_{i, j}^t$ represents the value, at time $t$, of cell $(i, j)$ of the generic variable $u^{}$. This approximation is justified numerically, as sketched above, and also because the scales of the biomass patterns are quite different (see Figure \ref{fig1}): by defining $h = 1$, the scale in the model is arbitrary and the discrete laplacian measures the influence of neighbor cells on the actual one. This idea is similar to those used in CA techniques in modeling several systems dynamics %{\cite{Tcainmath}}. 
  An important question about this redefinition of the discrete laplacian term is if the assumptions made avobe affect the performance of the model's emergent spatial patterns. The answer is in the negative sense, as can be seen in Figures \ref{manchas}, \ref{laberinto} and \ref{huecos}.
  
  {\noindent}Thus, Eq. (\ref{hard1}) and (\ref{hard2}) can be rewriten in
  their discrete version both temporal, using Euler's method, \ and spatial
  as:
%\end{flushleft}

\begin{equation} 
n^{t + 1}_{i, j} = \left( \frac{\gamma w^t_{i, j}}{1 + \sigma w^t_{i, j}}
   n^t_{i, j} - {n^t_{i, j}}^2 - \mu n^t_{i, j} + \nabla^2 n^t  \right) \Delta t
   + n^t_{i, j} \label{ec-hardenberg1}
\end{equation}

\[ w^{t + 1}_{i, j} = \left( p - (1 - \rho n) w^t_{i, j} - n {w^t_{i, j}}^2 +
   \delta \nabla^2 (w^t - \alpha n^t) + v \frac{\partial (w^t - \beta
   n^t)}{\partial x}  \right) \Delta t + w^t_{i, j} . \label{h2} \]
where
\[ \nabla^2 n^t = n^t_{i, j + 1} + n^t_{i, j - 1} + n^t_{i + 1, j} + n^t_{i -
   1, j} - 4 n^t_{i, j} \],

\begin{equation} 
  \nabla^2 (w^t - \alpha n^t) = w^t_{i, j + 1} + w^t_{i, j - 1} + w^t_{i + 1,
   j} + w^t_{i - 1, j} - 4 w^t_{i, j} - \alpha \nabla^2 n^t  
\end{equation}
 
and
\begin{equation}
  \frac{\partial (w^t - \beta n^t)}{\partial x} = \frac{w^t_{i + 1, j} +
  w^t_{i - 1, j} - \beta (n^t_{i + 1, j} + n^t_{i - 1, j})}{2}  \label{runnof}
\end{equation}

\subsection{Non flat regions}\label{nonflat}

%\begin{flushleft}
  {\noindent}In the search of a more complete model of arid and sem-arid
  ecosystems, a simplification of non flat regions, where there is no constant
  slope, is introduced because vegetation patterns disappear in Hardenberg's
  model when two different slope areas are used as the region where vegetation
  dynamics takes place.
  
  {\noindent}As a first attempt, a pair of joined plane regions, one of them
  with no null slope, are considered, opposite of what was done in
  Hardenberg's model where diferent slope regions are treated separately, as
  can be seen in Figure \ref{fig4}. This assumption is justified by field
  reports as in {\cite{agessler2000}}, so that any small or more or less
  regular region can be mapped by means of this two-region model. It is
  necessary to stablish adecuate boundary conditions in order to avoid the
  introduction of unphysical elements, and must reflect the natural dynamics
  of watter diffusion, as limitting resource.
  
  {\noindent}First of all, as a lattice of size $L \times L$ is used, the
  joint point within both regions is taken at a position $L / 2$ over
  direction $x$, so the parameter $\nu$ is non zero for $x \leq L / 2$ and
  equal to zero after this limit, for all the cells in the grid. In the
  extreme left, where hill begins, and in the extreme right, where the flat
  region ends, null and null flux border conditions are considered
  respectively, both for biomass and watter densities. In the direction $y$,
  where no slope effects are considered, periodicall border conditions are
  taken.
  
  {\noindent}In the joint region the border conditions can not be arbitrarily
  selected. The watter runnoff ends in a zero slope region increasing the
  humidity available for biomass growth there and, then, it begins to spread.
  Thus, in order to take account for this humidity increase, the average of
  watter runoff term in its discret form (\ref{runnof}), taken over each cell
  of the sloping region, is added to the correspondent first cell of the grid
  with zero slope. This average takes the form:
%\end{flushleft}

\[ < w_x > = \frac{v}{L} \sum_{j = 1}^{\frac{L}{2}} [w_{k, i + 1, j} - w_{k, i
   - 1, j} - \alpha (n_{k, i + 1, j} - n_{k, i - 1, j})], \]

%\begin{flushleft}
  {\noindent}Whith this construction, several runs were performed in order to
  reproduce spatial veggetation patterns in both regions, and the results can
  be observed in Figure \ref{sisalio}. The patterns shows certain similarity
  with those in real pictures and in the original results of Hardenberg's
  model (Figures \ref{originalH} and \ref{fig4}). Although these ``hybrid''
  patterns are a good aproximation to the real ones, they are still quite
  regular and its shape can be improved using the idea of a non constant
  difussion coefficient, as was done in some texture synthesis works
  {\cite{awitkin1991,asanderson2006}}, applied to water runoff term in Eq.
  (\ref{hard2}).
  
  {\noindent}In the last term of Eq. (\ref{hard2}), the one associated with
  the partial derivative, the constant factor $\nu$ is now considered as an
  explicit spatial dependant function wich takes the form:
%\end{flushleft}

\begin{equation}
  \nu (x) = \frac{1}{2} \nu \left[ \tanh \left( \frac{20 x}{l} - 10 \right) +
  1 \right] \label{nudex}
\end{equation}

%\begin{flushleft}
  {\noindent}where parameter $\nu$ is taken as in Eq. (\ref{hard2}) and $l$ is
  the number of unitary cells in lattice lenght $L$. As $\nu$ represents the
  watter runnoff velocity in a slope, this new function represents a region
  where slope is slightly increased from $0$ to a maximum value $\nu (x) =
  \nu$ as the position in the direction $x$ is increased. In doing so, it is
  clear that the joint conditions, explained above, are now unnecessary.
  
  {\noindent}In this aproach, zero flux border conditions are considered for
  the left and right extremes over $x$ direction and periodical border
  conditions over direction $y$. The resulting patterns are shown in Figure
  \ref{sinu}, and can be observed a significant improvement in the similarity
  with real patterns (Figure \ref{originalH}).
%\end{flushleft}

\section{Results}\label{resultados}

%\begin{flushleft}
  {\noindent}In order to know the possible correlations between local biomass
  density, humidity and the control parameter $p$ -that defines the model
  dynamic behavior- in this section vegetation density distributions for some
  fixed values of $p$ and the spatial patterns obtained in a non flat region
  with both aproaches explained in Section \ref{nonflat} are shown.
  
\begin{figure}[H]
  \centering
    \includegraphics[width=6in]{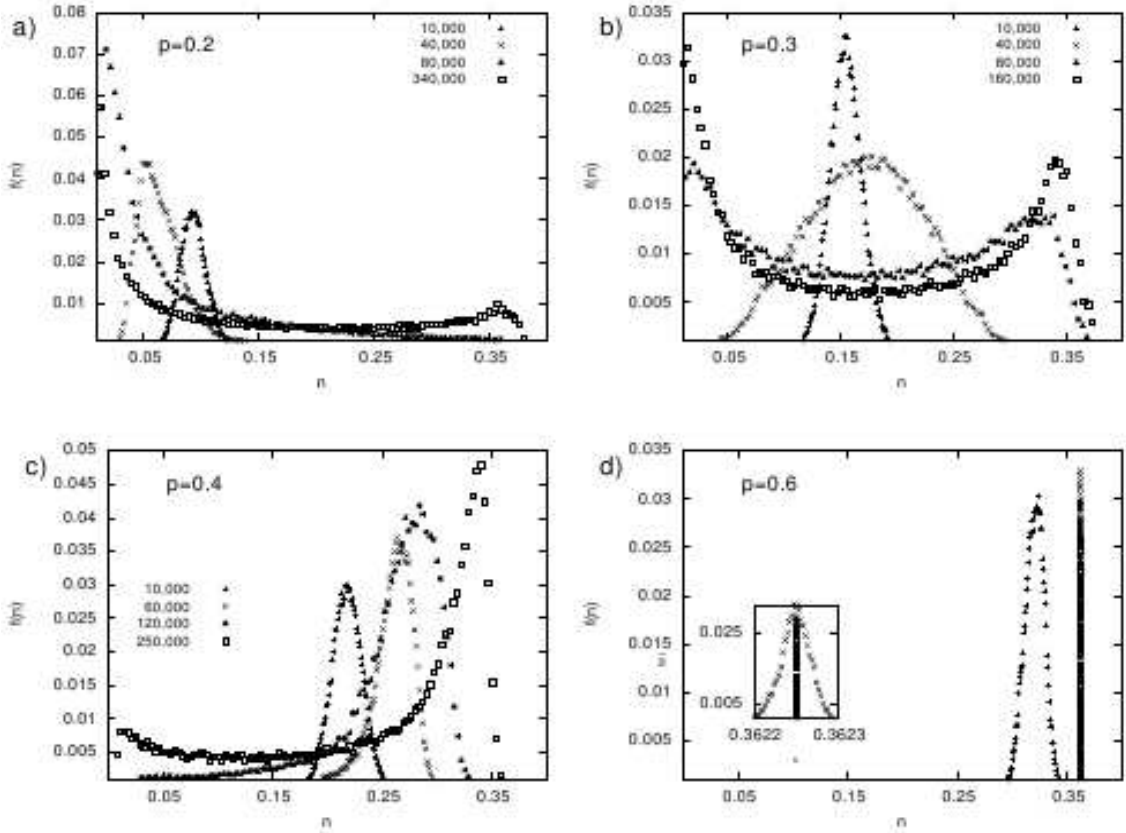}
    \caption{Evolution and stable statistical distributions of biomass density for some characteristic values of $p$. The fixed points for each $p$ value are: a) $n^{\star} =$ 0.05654; b) $n^{\star} =$0.17379; c) $n^{\star} =$0.26110 and d) $n^{\star} =$0.36225. Observe that biomass density distributions begins around fixed points values in all cases (marked with $+$), and that there exists a stable characteristic value for biomass density above $n=0.35$.}
  \label{densHard}
\end{figure}

  {\noindent}Consistence in a possible predictibility criterion implies that
  the temporal evolution of the variables leads them toward its fixed point
  values. Statistical distribution of biomass density, evaluated on all cells
  of the lattice for some $p$ values, are shown in Figure \ref{densHard} and
  its observed that its behavior depends on the $p$ parameter value.
%\end{flushleft}

%\begin{flushleft}
  {\noindent}The stable statistical distribution curve of biomass density
  (with small square points in all images of Figure \ref{densHard}) changes,
  when $p$ does, from predominant high values for small $n$ (see Figure
  \ref{densHard}a) with a small protuberance near of $n \simeq 0.36$ (for $p =
  0.2)$, in passing by predominant small and high $n \simeq 0.36$ values of
  density (Figure \ref{densHard}b), toward a delta distribution for high
  values of $p$ (Figures \ref{densHard}c and \ref{densHard}d). In all cases,
  statistical distributions begins at values near to its respective fixed
  point and, as can be seen in the inset in Figure 7d, only one tends to this
  value. In the other cases (7a, 7b and 7c) the distribution moves away from
  the fixed point.
  
\begin{figure}[H]
  \centering
  \includegraphics[width=1.5in]{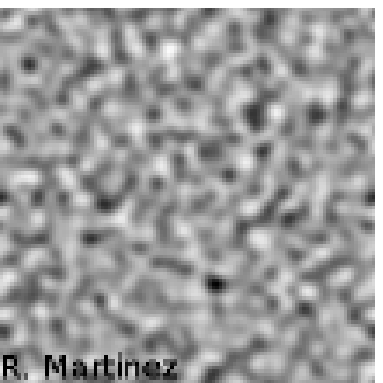}
  \includegraphics[width=1.5in]{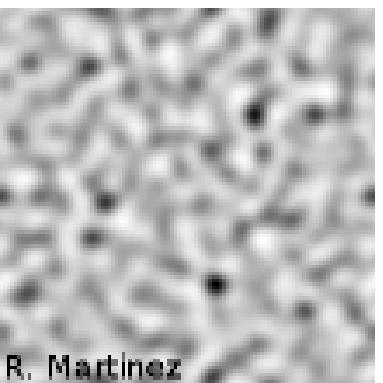}
  \includegraphics[width=1.5in]{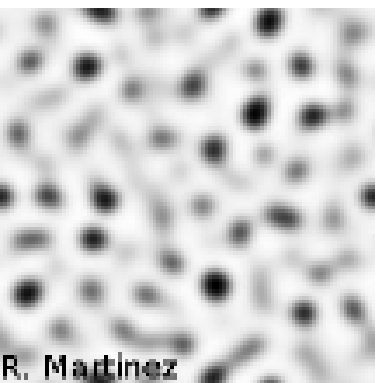}
  \includegraphics[width=1.5in]{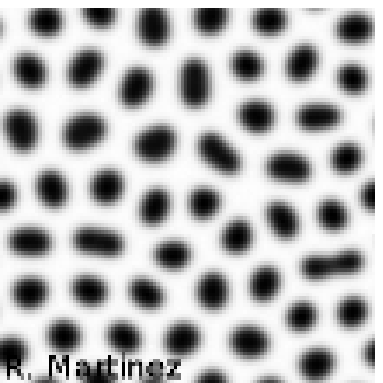}
  \caption{\label{manchas}Hardenberg's patterns of biomass: {\em{spots}} for $p = 0.2$. Here the cells with very low values  if density dominates the space.}
\end{figure}

\begin{figure}[H]
  \begin{center}
    \includegraphics{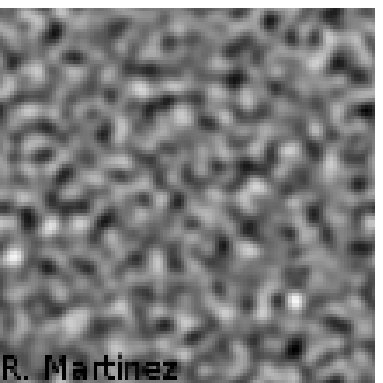}
    \includegraphics{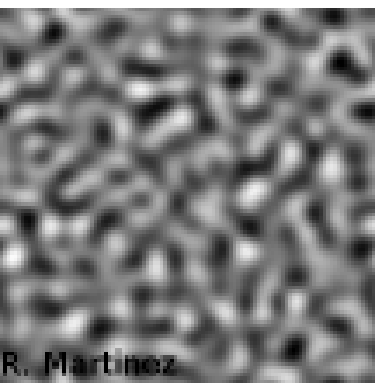}
    \includegraphics{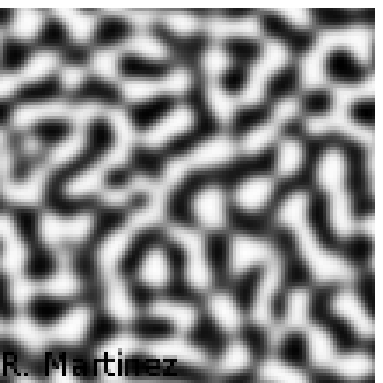}
    \includegraphics{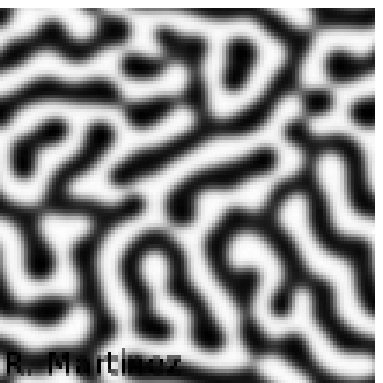}
    \caption{\label{laberinto}Evolution of
    biomass patterns toward a {\em{labyrinth}} for \ $p \text{} = 0.3$, the sites of low density are spacially distributed in analog to high density cells. This fact corresponds to a pair of stable maximum values in density distribution (see Fig. \ref{densHard} b).}
  \end{center}
\end{figure}

\begin{figure}[H]
  \includegraphics{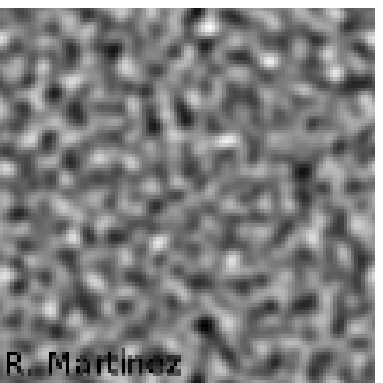}
  \includegraphics{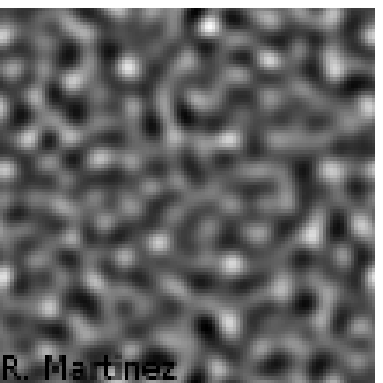}
  \includegraphics{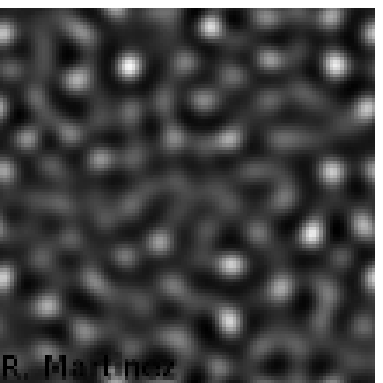}
  \includegraphics{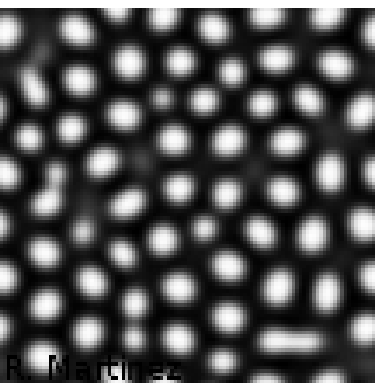}
  \caption{\label{huecos}Evolution of Hardenberg's biomass: {\em{holes}} for $p = 0.4$, where high values of density dominates over low density regions.}
\end{figure}

\begin{figure}[H]
  \centering
  \includegraphics{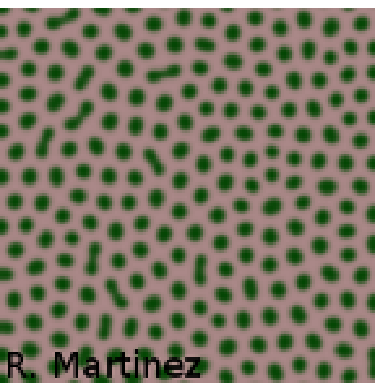}
  \includegraphics{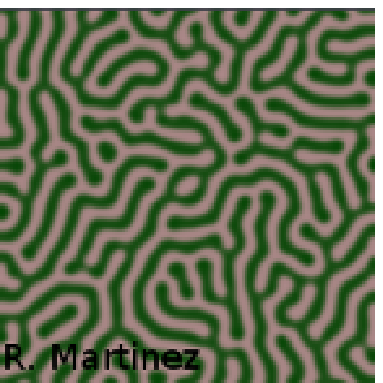}
  \includegraphics{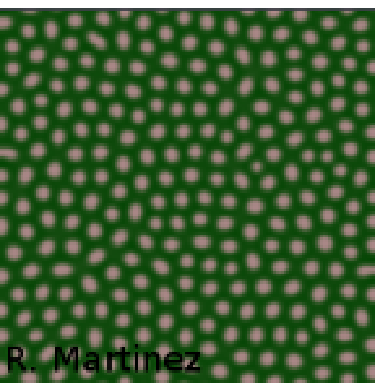}\caption{\label{flat512patterns}Patterns
  obtained from a $512 \times 512$ grid with $\nu = 0$.}
\end{figure}

  {\noindent}Figures \ref{manchas}, \ref{laberinto}, \ref{huecos} and
  \ref{flat512patterns} shows the stable emergent patterns of biomass spatial
  distribution. This patterns are in accordance with the real ones (Figure
  \ref{originalH}), and with those obtained by Hardenberg (Figure \ref{fig4}),
  which shows that a discrete version like that proposed in Section
  \ref{discrete}, where the evolution local of the variable $n$ depends on its
  neighbor values, is a good aproximation in order to qualitatively represent
  vegetation pattern formation process. In fact, models with evolution rules
  based in neighbor cell values, such as the CA models {\cite{tSmith94}} and
  the above mentioned model of Shnerb {\cite{amanor2008}}, has shown
  acceptable results in modeling various dynamical systems
  {\cite{lsole,aiwasa91,abascompteaguja}}.
%\end{flushleft}

\begin{figure}[H]
  \centering
  \includegraphics{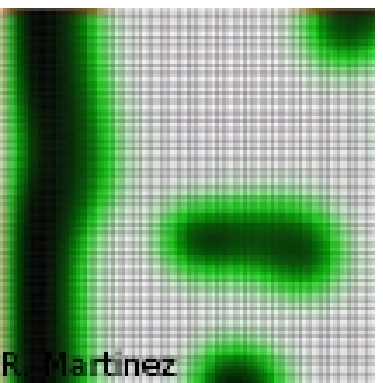}
  \includegraphics{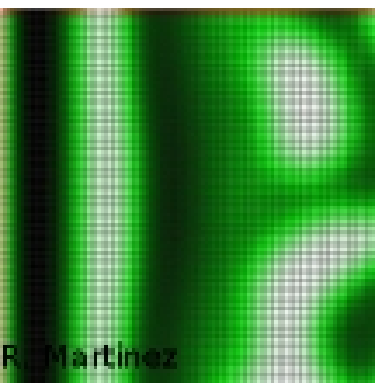}
  \includegraphics{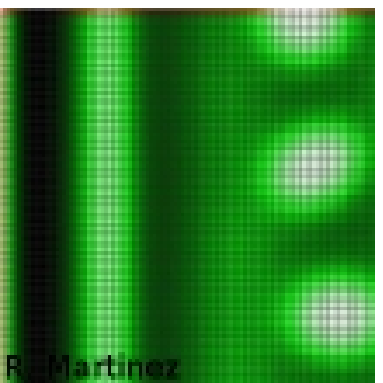}
  \caption{\label{sisalio}Left to right, p=0.2, 0.3, 0.4, same values of $p$ that in figures \ref{manchas}, \ref{laberinto} y \ref{huecos}. The patterns are combination of patterns: patterns of inclined surface (left in all grids) \ joined with patterns of horizontal plane (right side on all grids).}
\end{figure}

%\begin{flushleft}
  {\noindent}Vegetation patterns obtained v\'{\i}a ``joint'' conditions and
  the smoothing function $\nu (x)$ are presented in Figures \ref{sisalio} and
  \ref{sinu}, respectively. Observe that in mixed patterns in the former
  (Figure \ref{sisalio}) the association between banded patterns and those
  observed inThis null slope regions is near to the middle of the lattice,
  and, in both figures, this association is made via an unfinished banded
  pattern merged with those in zero slope region. Note that, while those
  presented in Figure \ref{sisalio} are in a certain way ``regular'' and de
  band is almost straight, patterns in Figure \ref{sinu} are closer to real
  vegetation patterns.
%\end{flushleft}

\begin{figure}[H]
  \centering
  \includegraphics{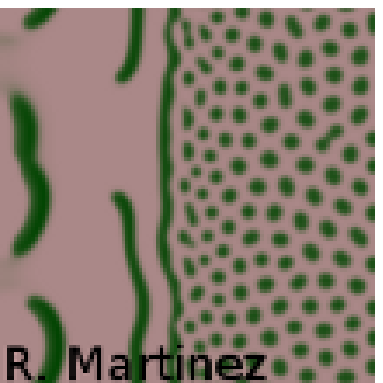}
  \includegraphics{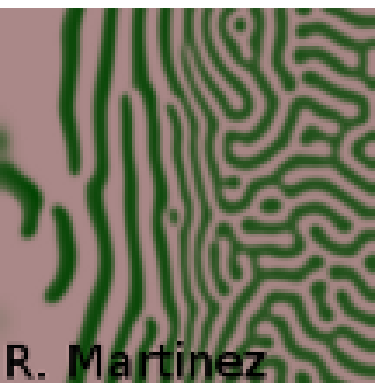}
  \includegraphics{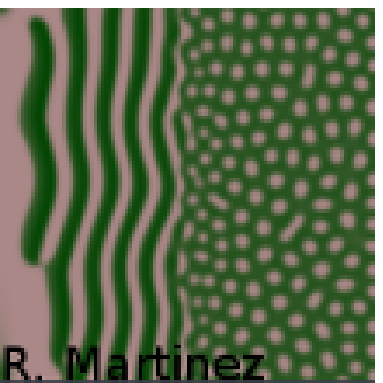}
  \caption{\label{sinu}In a similar grid than Figure \ref{flat512patterns} using the smoothing function $\nu (x)$ in Eq. (\ref{nudex}). The pictures above were rotated $180^{\circ}$ to show the correspondence with those in Figure \ref{sisalio}}
\end{figure}

\section{Discussion}

  The behaviour of biomass density patterns and the associated
  statistical distribution must be sufficient as criterion to determine
  biomass density from images and, eventually, they could be used in
  determination of zones in dessertification risk. Very small values of $n$
  must correspond to null plants regions as can viewed from Figure
  \ref{densHard}a. Surprisingly, although the dynamical behaviour tends to the
  fixed points (phase portraits in Figure \ref{puntosfijosH}) biomass density
  distributions moves away from fixed point value, except for large $p$ values
  (Figure \ref{densHard}d). This is so because spatial patterns appear when a
  stationary stable state in absence of diffusion -in other words a fixed
  point- becomes unstable under diffusion, which is surprising because usually
  when a substance begins to spread, concentration gradients of it decreases
  in time, which leads to the disappearance of spatial configurations
  {\cite{morphoTuring,abarrio2010}}. In this cases, diffusion mechanism is
  responsible of pattern formation and of the behaviour observed in the
  statistical distibutions \ref{densHard}a, \ref{densHard}b and
  \ref{densHard}c.
  
  {\noindent}Also, note that in Figure \ref{densHard} no L\'evy-like regime
  emerges in the statistical distributions of biomass density, although in
  Figure \ref{densHard}a, which corresponds to low $p$ values, a sort of
  broken L\'evi-like regime, at high values of biomass density, is shown as
  was seen in the last image in Figure \ref{fig3} for the patch size
  distribution of the model at {\cite{amanor2008}}.
  
  {\noindent}Note that the pattern formation process continues even when the
  discretization criterion, and its consequent loss of scale, is applied to
  difussion term in Eqs. (\ref{hard1}) and (\ref{hard2}) as was explained in
  section \ref{discrete}. This simplification, chosed due to the need of
  obtain the spatial patterns from a randomized initial condition and due to
  the diferent scales of vegetation spatial patterns, shows a more realistic
  representation of spatial configurations presented as an emergent behaviour
  of vegetation in arid and semi-arid areas, as can be seen in Figure
  \ref{originalH}.
  
  {\noindent}Furthermore, the ``joint'' conditions and spatial explicit
  dependance of watter runoff velocity $\nu$, in Eq. \ref{hard2}, proposed in
  Section \ref{nonflat}, allows the creation of patterns that mix spatial
  shapes observed in nature (Figures \ref{originalH} and \ref{fig4}), which
  can not be done with Hardenberg's model alone, as the patterns disappear.
  Also, this construction brokes the linear behaviour of banded patterns
  obtained using Hardenberg's model alone (Figure \ref{fig4}) making them
  closer to the real ones. Thus, the variable slope of the landscape could be
  another posible explanation to the irregularity of natural patterns (Figure
  \ref{originalH}) along with seed dispersal process, as was stablished in
  {\cite{athompson2008}}.
  
  {\noindent}Nevertheless, more general models can involve
  cooperation-competition events between different vegetal species doing
  modifications in the parameters $\rho$, associated with the external
  morphology of the plant in terms of shadow capacity, $\alpha$ with the
  absorption capacity -that diminishes the local diffusion of humidity- and
  $\beta$ that can be related with both the local absorption capacity and with
  a physical barrier that dumps the water flux. Also, the resulting patterns,
  and its associated vegetation density, could be used as food resource maps
  where foragers movement takes place, in a similar way as in {\cite{aboyer}}.

\section*{Acknowledgments}
We are grateful for the facilities provided by the Laboratorio Nacional de Supercómputo (LNS) del Sureste de México to obtain these results.

%\section{References}
%\label{S:6}
%\begin{thebibliography}
\bibliographystyle{unsrt}
%\bibliography{references}
%\end{thebibliography}

\end{document}